\documentclass[aps,showpacs,
amssymb,endfloats]{revtex4}
\usepackage{amsmath}
\usepackage{bm,amsfonts,graphicx}
\begin{document}
\title {Beam splitting and Hong-Ou-Mandel interference for stored light}
\author{A. Raczy\'nski}
\email{raczyn@phys.umk.pl}
\author{J. Zaremba}
\affiliation{Instytut Fizyki, Uniwersytet Miko\l aja Kopernika,
ulica Grudzi\c{a}dzka 5, 87-100 Toru\'n, Poland,}
\author{S. Zieli\'nska-Kaniasty}
\affiliation{Instytut Matematyki i Fizyki, Akademia
Techniczno-Rolnicza, Aleja Prof. S. Kaliskiego 7, 85-796
Bydgoszcz, Poland.}
\begin{abstract}
Storing and release of a quantum light pulse in a medium of atoms in the tripod configuration are
studied. Two complementary sets of control fields are defined, which lead to independent and complete
photon release at two stages. The system constitutes a new kind of a flexible beam splitter in which
the input and output ports concern photons of the same direction but well separated in time.
A new version of Hong-Ou-Mandel interference is discussed.
\end{abstract}
\maketitle
\newpage
\section{Introduction}
Light propagation and storing in laser driven atomic media have recently become a subject of
numerous studies, stimulated both by a fundamental character of the discussed problems and
by possible applications. A generic system for such studies is a medium of three-level atoms
interacting with two laser beams (signal and control) in the $\Lambda$ configuration. The
essence of the process of light storing is a coherent mapping of the signal pulse into an
atomic excitation, described by an atomic coherence, due to a switch-off of a control laser
beam. By switching the control field again one restores the trapped pulse, preserving all
the phase relations. This means in fact first writing down an information conveyed by the
photons and later reading it back. The whole process can be effectively interpreted in terms
of a joint medium+field excitation called dark state polariton. For recent reviews on light
storage see, e.g., Refs \cite{lukin,andre,imamoglu}.

Admitting additional fields and thus additional active atomic states enriches the dynamics
of the process and gives new possibilities of its control. In particular by adding another
upper state one obtains a double $\Lambda$ system \cite{hemmer} in which it is possible to
make the medium transparent simultaneously for two signal pulses or to split a signal pulse
into two ones of different frequencies \cite{my1,xiong}. Channelization of information is
also possible in the case of inverted Y systems \cite{joshi1}.  Coupling the $\Lambda$
system with a side level makes it possible to release a pulse of a frequency different from
the original one or to temporarily prevent the signal from being released \cite{my2}.
Recently the dynamics of light propagation in 5-level atomic systems in the $M$
configuration has also been discussed \cite{pengbo}.

A special example is the so-called tripod atomic system in which three low-lying levels are
coupled with an upper level by a signal field and two control ones. By a manipulation on
control fields one can steer the propagation, storing and release of the signal, which may
be stored in the form of two atomic coherences. Light propagation and slowdown in such a
medium have been studied in Refs \cite{pasp,maz} while the problem of light storing has been
discussed in detail in our recent paper \cite{my}. An attempt of releasing the trapped pulse
by an arbitrary combination of two control fields results in general in splitting it into
two parts one of which was leaving the sample while the other one remained trapped and could
be released by switching on a new set of control fields. The evolution of such a system
could be effectively described in terms of a couple of polaritons.

The previously used classical description of dividing the pulse into two or more parts may
be essentially enriched by treating the pulse quantum-mechanically and by asking new
questions about the photon statistics, especially important in the case of nonclassical
light states, e.g., Fock states with a small photon number. Such an approach is also
necessary in the consideration of quantum information storage in an atomic memory. Wang {\em
et al.} \cite{yel} demonstrated a possibility of time splitting of a single photon by first
storing it in one coherence, than by a sequence of transferring a part of the excitation
into the other coherence and of a photon release from the latter. An additional transfer
procedure based on a fractional adiabatic population transfer (F-STIRAP) had to be performed
before each release procedure.

In this paper  we investigate light propagation and storage of a quantum signal field in a
medium of atoms in the tripod configuration. We discuss mapping of a quantum wave packet
into two atomic coherences which means in particular a channelization of a single photon. In
our approach, being more general than that of Ref. \cite{yel}, at the release stage the two
control fields may be twice switched on and off simultaneously. If the control pulses are
complementary, which means that their amplitude and phase relations are properly chosen, the
whole of the stored signal is divided into two parts, which are released at two separated
time instants. In the case of a single trapped photon this leads to its time-entangled
state, but without a need of an additional transfer stage of Ref. \cite{yel}. We discuss in
detail the case in which two time-separated photons are trapped at the same place of the
sample due to an action of two complementary sets of control pulses. They are later released
by a different pair of sets of such pulses. We examine the possibility of photon
coalescence, i.e. a release of both photons at the same release stage. By changing
localizations and shapes of the stored photon wavepackets and the details of the control
fields we obtain an analogue of the Hong-Ou-Mandel interferometer
\cite{hong,grang,knight,olindo} working on stored light. In contradistinction to the
original version of such an interferometer we have to do with photons propagating always in
the same spatial direction but well resolved in time.

\section{Quantum polaritons in the tripod case}

In order to investigate quantum effects in light storing and propagation both the signal and
the medium have to be described quantum-mechanically. Some formulae of this section,
pertaining to quantum operators, resemble those presented in our previous paper \cite{my}
for classical pulse. However, they are rederived below not only to make the presentation
complete but also because they will serve to a discussion of physical effects connected with
a quantum field, in particular with single photon states. In particular polaritons which in
the present formulation are quantum-mechanical operators will play a central role in
describing photon correlation effects.

Consider a one-dimensional medium of atoms in the tripod configuration (Fig. \ref{fig1})
including an upper state $a$ and three stable lower states $b$ (initial), $c$ and $d$. The
sample is irradiated with three collinear laser beams. The quantized signal field couples
the states $b$ and $a$ and is written as:
\begin{equation}
\epsilon(z,t)=\epsilon^{(+)}(z,t)+\epsilon^{(-)}(z,t)=
\Sigma_{k}g_{k}a_{k}\exp[i(kz-\omega
t)]\exp[-i(k_{ab}z-\omega_{ab}t)] +h.c.,
\end{equation}
where $a_{k}$ is the annihilation operator of a photon with wave number $k$,
$g_{k}=\sqrt{\frac{\hbar\omega}{2\epsilon_{0}V}}$, $V$ is the quantization volume,
$\epsilon_{0}$ - the vacuum electric permittivity and
$k_{\alpha\beta}=\omega_{\alpha\beta}/c= (E_{\alpha}-E_{\beta})/(\hbar c)$,
$\alpha,\beta=a,b,c,d$. The two control fields $\epsilon_{2,3}(t)
\cos(k_{2,3}z-\omega_{2,3}t)$ are treated classically and are supposed to be strong enough
for the propagation effects to be neglected; they couple respectively the states $c$ and $d$
with $a$. The medium excitation is described in the formalism of the second quantization by
the flip operators $\sigma_{\alpha\beta}=|\alpha><\beta| \exp(ik_{\alpha\beta}z)$,
$k_{\alpha\beta}$\ being the corresponding wave vector. The medium is treated in a
continuous way; this is why $\frac{N}{L}\int dz$ in the above equation has replaced the sum
over $N$ atoms ($L$ is the length of the sample).

The interaction hamiltonian in the rotating-wave approximation
reads
\begin{eqnarray}
H=\frac{N}{L}\int
dz(-d_{ab}\sigma_{ab}\exp[-i(k_{ab}z-\omega_{ab}t)]\Sigma
a_{k}g_{k}\exp[i(kz-\omega t)] + h.c.\nonumber\\
-\frac{1}{2}d_{ac}\sigma_{ac}\exp[-i(k_{ac}z-\omega_{ac}t)]
\epsilon_{2}(t)\exp[i(k_{2}z-\omega_{2}t)]+h.c.\\
-\frac{1}{2}d_{ad}\sigma_{ad}\exp[-i(k_{ad}z-\omega_{ad}t)]
\epsilon_{3}(t)\exp[i(k_{3}z-\omega_{3}t)]+h.c.),\nonumber
\end{eqnarray}
where $d_{\alpha\beta}$ are the dipole moment's matrix elements, assumed for simplicity to
be real and positive.

In our analysis in terms of polaritons it is enough to restrict oneself to the deterministic
part of the Heisenberg evolution (relaxation- and noise-free). The equations of motion for
the flip and signal field operators in the first-order approximation with respect to the
signal field, in resonance conditions read
\begin{eqnarray}
&i\dot{\sigma}_{ba}(z,t)=-\frac{d_{ab}}{\hbar}\epsilon^{(+)}(z,t)
-\Omega_{2}(t)\sigma_{bc}(z,t)-\Omega_{3}(t)\sigma_{bd}(z,t),\nonumber\\
&i\dot{\sigma}_{bc}(z,t)=-\Omega_{2}^{*}(t)\sigma_{ba}(z,t),\\
&i\dot{\sigma}_{bd}(z,t)=-\Omega_{3}^{*}(t)\sigma_{ba}(z,t),\nonumber\\
&(\frac{\partial}{\partial t}+c\frac{\partial}{\partial
z})\epsilon^{(+)}(z,t)=\frac{iN}{\hbar}g^{2}
d_{ba}\sigma_{ba}(z,t),\nonumber
\end{eqnarray}
where $g$ is the value of $g_{k}$
for the central field frequency of the signal field, and
$\Omega_{2}(t)=d_{ac}\epsilon_{2}(t)/(2\hbar)$ and
$\Omega_{3}(t)=d_{ad}\epsilon_{3}(t)/(2\hbar)$ are the Rabi
frequencies for the control fields.

It is convenient to express the solutions of Eqs (3) in terms of two field+medium
excitations called polaritons: $\Psi(z,t)$ (the dark-state polariton) and $Z(z,t)$
\begin{eqnarray}
&\Psi(z,t)=\exp(-i\chi)\{\epsilon^{(+)}(z,t) \cos\theta-\frac{\hbar\kappa}{d_{ab}}
\sin\theta[\exp(i\chi_{2})\cos\phi\sigma_{bc}(z,t)+
\exp(i\chi_{3})\sin\phi\sigma_{bd}(z,t)]\},
\nonumber\\
&Z(z,t)=\frac{\hbar\kappa}{d_{ab}} [\exp(i\chi_{2})\sin\phi\sigma_{bc}(z,t)-\exp(i\chi_{3})
\cos\phi\sigma_{bd}(z,t)],
\end{eqnarray}
where the mixing angles $\theta$ and $\phi$ are defined as $\tan\theta= \kappa/\Omega$,
$\tan\phi=|\Omega_{3}/\Omega_{2}|$ with $\chi_{2}=\arg(\Omega_{2})$,
$\chi_{3}=\arg(\Omega_{3})$, $\Omega=\sqrt{|\Omega_{2}|^{2}+|\Omega_{3}|^{2}}$,
$\kappa^{2}=Ng^{2}|d_{ab}|^{]2}/\hbar^{2}$; $\chi$ satisfies the equation
$\dot{\chi}=\sin^{2}\theta\dot{\chi}_{2}$. For simplicity we assume that
$\dot{\chi}_{2}=\dot{\chi}_{3}$. The field and medium components of the polariton $\Psi$ are
\begin{eqnarray}
&\epsilon^{(+)}=\exp(i\chi)\Psi\cos\theta,\\
&\exp(i\chi_{2})\cos\phi\sigma_{bc}+\exp(i\chi_{3})\sin\phi\sigma_{bd}=
-\frac{d_{ab}}{\hbar\kappa}\exp(i\chi)\Psi\sin\theta.
\end{eqnarray}

The evolution equations for the two polaritons read
\begin{eqnarray}
&(\frac{\partial}{\partial t}+c\cos^{2}\theta
\frac{\partial}{\partial z})\Psi=
\exp(-i\chi)\sin\theta\dot{\phi} Z,\nonumber\\
&\frac{\partial}{\partial t}Z=i\dot{\chi}_{2}Z- \exp(i\chi)\sin\theta\dot{\phi}\Psi.
\end{eqnarray}

The operators $\Psi(z,t)/(g\sqrt{L})$ and $Z(z,t)/(g\sqrt{L})$ fulfil typical bosonic
commutation relations in the first-order approximation, in which $\sigma_{bb}=1$,
$\sigma_{cc}=\sigma_{dd}=\sigma_{cd}=0$,
($[\Psi(z),\Psi^{\dagger}(z')]=[Z(z),Z^{\dagger}(z')]= g^{2}L \delta(z-z')$,
$[\Psi(z),Z^{\dagger}(z')]=0$), so those operators may be considered annihilation operators
of the joint field+medium excitation. The essence of the approach used in this paper will be
treating a medium excitation due to the stored photon as a superposition of two excitations,
properly suited to the pulse releasing control fields (see Eq. (9) below).

If the control fields are proportional, i.e. $\dot{\phi}=0$ and
$\dot{\chi}_{2}=\dot{\chi}_{3}=0$
the evolution equations for the polaritons are decoupled.
The polariton $\Psi$ propagates through the medium
without changing its "shape", i.e. the solution is
\begin{equation}
\Psi(z,t)=\Psi(z-c\int_{0}^{t}\cos^{2}\theta(t')dt',t=0),
\end{equation}
while the polariton $Z=const$ keeps both its shape and localization.

The above equations constitute a generalization of the approach of Fleischhauer et al.
\cite{fl} for a four-level system and at the same time of our previous results \cite{my},
here formulated in the language of a quantum signal field. Due to such an approach we will
be able to discuss the effect of the process in terms of photons rather than in the language
of splitting a classical pulse.

If at the beginning the control fields corresponding to a given constant mixing angle
$\phi^{0}$ and given constant $\chi_{2}^{0}$ and $\chi_{3}^{0}$ are strong enough
($\theta\approx0$), the medium is transparent and an incoming photon enters the medium.
Switching the control field off ($\theta=\frac{\pi}{2}$) results in the medium becoming
opaque and the photon becomes trapped in a coherent superposition of the atomic coherences.
The propagation and storing are described in terms of the polaritons $\Psi^{0}$ and $Z^{0}$.
The former polariton has evolved from being almost purely electromagnetic to become purely
atomic while the latter one has not taken part in the evolution. The probability amplitudes
that the photon has actually been trapped in the coherence $\sigma_{bc}$ and $\sigma_{bd}$
are respectively $\exp(i\chi_{2}^{0})\cos\phi^{0}$ and $\exp(i\chi_{3}^{0})\sin\phi^{0}$. If
now the control fields of the same mixing angle and the same phases are switched on again
the photon will be released with certainty in the same state as the initial one. This is
{\em mutatis mutandis} a repetition of the case of a usual $\Lambda$ system discussed
before.

If the control fields to be used at the release stage are characterized by the angles
$\phi^{1}$, $\chi_{2}^{1}$ and $\chi_{3}^{1}$ it is proper to describe the further evolution
in terms of the corresponding new polaritons $\Psi^{1}$ and $Z^{1}$. The latter pair of
polaritons is expressed in terms of the former ones as

\begin{eqnarray}
\Psi^{1}=\{\cos\phi^{1}\cos\phi^{0}\exp[i(\chi_{2}^{1}-\chi_{2}^{0})] +
\sin\phi^{1}\sin\phi^{0}\exp[i(\chi_{3}^{1}-\chi_{3}^{0})]\}\Psi^{0} +\nonumber\\
\{\cos\phi^{1}\sin\phi^{0}\exp[i(\chi_{2}^{1}-\chi_{2}^{0})]
-\sin\phi^{1}\cos\phi^{0}\exp[i(\chi_{3}^{1}-\chi_{3}^{0})]\}Z^{0},\\
Z^{1}=\{\sin\phi^{1}\cos\phi^{0}\exp[i(\chi_{2}^{1}-\chi_{2}^{0})] -
\cos\phi^{1}\sin\phi^{0}\exp[i(\chi_{3}^{1}-\chi_{3}^{0})]\}\Psi^{0} +\nonumber\\
\{\sin\phi^{1}\sin\phi^{0}\exp[i(\chi_{2}^{1}-\chi_{2}^{0})]
+\cos\phi^{1}\cos\phi^{0}\exp[i(\chi_{3}^{1}-\chi_{3}^{0})]\}Z^{0}\nonumber.
\end{eqnarray}

The polariton $\Psi^{1}$ will be released at turn into a photon, while the $Z^{1}$ will
remain trapped. It follows from the form of the polaritons (Eq. (4)) that applying another
pair of the control fields with $\phi^{2}=\frac{\pi}{2}-\phi^{1}$, $\chi_{2}^{2}=
\chi_{2}^{1}+\pi$ and $\chi_{3}^{2}=\chi_{3}^{1}$ leads to an exchange of the role of the
polaritons: $Z^{1}$ becomes $\Psi^{2}$ which will be released and turn into a photon. The
sets of control fields marked with (1) and(2) can be considered complementary in the sense
that their subsequent application implies a certain photon release.

The above results are much more general than those of Ref. \cite{yel} in which the authors
proposed a time splitting of a photon by first storing the pulse in one coherence
$\sigma_{bc}$ (here this corresponds to $\phi^{0}=0$), than by pumping a part of the
excitation into the other coherence $\sigma_{bd}$ and finally by a pulse release from the
latter coherence (here $\phi^{1}=\frac{\pi}{2}$). The pumping procedure (F-STIRAP) was to be
separated from the release stages which required using light pulses transverse to both
signal and control fields; the latter should be realized in ultracold gases. Here, due to
using simultaneously pairs of control fields of the same shape pumping and releasing stages
coincide.

\section{Beam splitter and Hong-Ou-Mandel - type interferometer}

The above-described behavior of the atom+field allows one to use the medium and the system
of control fields as an effective and flexible beam splitter working on stored light. The
technical difference between the usual device and that of ours is that incoming photons may
arrive from only one direction but at one or two different time instants. Also the released
photons have the same direction but are separated in time. By changing the mixing angle
$\phi^{1}$ and the phases $\chi_{2}^{0,1}$ and $\chi_{3}^{0,1}$ of the control fields one
can smoothly regulate the amplitudes and phases of the pulse components released at the two
stages, which would correspond to changing the transmission and reflection rate of a usual
beam splitter.

The operation of photon storing can be performed at two stages (input ports). First we trap
a first portion of incoming photons by switching on and off a pair of control fields
characterized by the parameters $\phi^{0}$,$\chi_{2}^{0}$ and $\chi_{3}^{0}$. After the this
part of the trapping operation has been finished we may trap a second portion of photons by
applying the control fields of the mixing angle $\frac{\pi}{2}-\phi^{0}$, $\chi_{2}^{0}+\pi$
and $\chi_{3}^{0})$. Note that the second trapping operation does not affect the coherences
due to the first one. Thus a photon from the first portion is described by the polariton
field $\Psi^{0}$ and that from the second portion - by $Z^{0}$. A special case is assumed in
which $\phi^{0}=0$ which means that photons have been first trapped in the coherence
$\sigma_{bc}$ and next  - in $\sigma_{bd}$.

The release operation consists also of two stages (output ports) connected with two pairs of
control fields separated in time, the first of which is characterized by the mixing angle
$\phi^{1}$ and $\chi_{2}^{1}$ and $\chi_{3}^{1}$ and the second by $\frac{\pi}{2}-\phi^{1}$,
$\chi_{2}^{1}+\pi$ and $\chi_{3}^{1}$. The output includes thus two portions of photons,
separated in time, and the photon release is complete.

In the case of trapping and releasing of two photons we have a new realization of the
Hong-Ou-Mandel effect. In its original formulation it concerns a 50-50 photon beam splitter
with exactly one incoming photon in each of the two input ports (incoming photon
directions). Due to interference phenomena the probability amplitudes for obtaining exactly
one photon in each of the two output ports (outgoing photon directions) cancel out.

In our realization of the input ports single photons, separated in time, are trapped, one in
each of the two superpositions of the atomic coherences, connected with complementary sets
of control fields. In the output ports photons are released, again at two stages, from two
other superpositions of the two coherences. The analogue of photon incoming from two
orthogonal directions is their storing at two storing stages. The analogues of reflection
and transmission on a traditional beam splitter are photon release at the first or second
release stage while the parameters characterizing the control fields and the atomic system
determine the analogues of the reflection and transmission coefficient and phase relations.
Using the above-mentioned transformations of polaritons at the storing stage and having in
mind that the incoming polaritons $\Psi^{0},Z^{0}$ (outgoing polaritons $\Psi^{1},Z^{1}$)
become almost purely electric field operators $\epsilon^{1},\epsilon^{2}$,
($\epsilon^{3},\epsilon^{4}$) for $t\rightarrow -\infty$ ($t\rightarrow \infty$) we may give
the net result for the transformation of the field operators

\begin{equation}
\left(\begin{array}{cc}R_{31}&R_{32}\\R_{41}&R_{42}\\ \end{array}\right) \left(
\begin{array}{c}\epsilon^{1}\\\epsilon^{2}\\ \end{array}\right)\rightarrow
\left(\begin{array}{c}\epsilon^{3}\\\epsilon^{4}\\ \end{array}\right).
\end{equation}
where
\begin{eqnarray}
R_{31}=\cos\phi^{1}\cos\phi^{0}\exp[i(\chi_{2}^{1}-\chi_{2}^{0})] +
\sin\phi^{1}\sin\phi^{0}\exp[i(\chi_{3}^{1}-\chi_{3}^{0})],\nonumber\\
R_{32}=\cos\phi^{1}\sin\phi^{0}\exp[i(\chi_{2}^{1}-\chi_{2}^{0})]
-\sin\phi^{1}\cos\phi^{0}\exp[i(\chi_{3}^{1}-\chi_{3}^{0})],\nonumber\\
R_{41}=\sin\phi^{1}\cos\phi^{0}\exp[i(\chi_{2}^{1}-\chi_{2}^{0})] -
\cos\phi^{1}\sin\phi^{0}\exp[i\chi_{3}^{1}-\chi_{3}^{0})],\nonumber\\
R_{42}=\sin\phi^{1}\sin\phi^{0}\exp[i(\chi_{2}^{1}-\chi_{2}^{0})] +
\cos\phi^{1}\cos\phi^{0}\exp[i\chi_{3}^{1}-\chi_{3}^{0})].
\end{eqnarray}

Let the respective wave packets of the stored excitation be $f_{1}(z)$ and $f_{2}(z)$, their
shape being identical with the shape of the initial photon wavepackets and their
localization depending on the time instants of the switch-off of the control fields. The
maximum value (unity) of the overlap of the packets $s\equiv \int f_{j}^{*}(z)f_{k}(z)dz$
corresponds to the situation in which two photons with wavepackets of the same shape have
been stored exactly at the same place inside the sample.

The field operators corresponding to the input ports are $\Psi^{0}(j)=\int f_{j}^{*}(z)
\Psi^{0}(z)dz$, $Z^{0}(j)=\int f_{j}^{*}(z) Z^{0}(z)dz$. The corresponding operators in the
output ports are $\Psi^{1}(j)=\int f_{j}^{*}(z) \Psi^{1}(z)dz$, $Z^{1}(j)=\int f_{j}^{*}(z)
Z^{1}(z)dz$.

The key commutation relations are
$[\Psi^{0}(j),\Psi{^{0}}^{\dagger}(j)]=[Z^{0}(j),{Z^{0}}^{\dagger}(j)]=g^{2}L$,
$[\Psi^{0}(1),\Psi{^{0}}^{\dagger}(2)]=[Z^{0}(1),{Z^{0}}^{\dagger}(2)]=g^{2}L s$, with
analogous relations for $\Psi^{1}$ and $Z^{1}$.

The quantum state of the medium after trapping has been accomplished is constructed as due
to a creation of two excitations due to complementary control fields and characterized by
two possibly different wave packets
\begin{equation}
|\zeta>=\frac{1}{g^{2}L} {\Psi^{0}}^{\dagger}(1)Z{^{0}}^{\dagger}(2)|0>,
\end{equation}
where the "vacuum" state $|0>$ means all atoms in the state $|b>$). The state corresponding
to two photons being released at the first stage (photon coalescence) is constructed as due
to a creation of two excitations corresponding to the first release stage characterized by
possibly different wave packets
\begin{equation}
|\zeta_{1}>=\frac{1}{g^{2}L}\frac{1}{\sqrt{1+|s|^{2}}}
{\Psi^{1}}^{\dagger}(1)\Psi{^{1}}^{\dagger}(2)|0>,\
\end{equation}
where a normalizing factor for the two-particle states has been included. Note that the
state corresponding to the photon coalescence at the second stage is constructed in an
analogous way, using the $Z$ operators instead of $\Psi$.

Calculating the projections with the use of the above mentioned commutation relations yields
the probability amplitude of photon coalescence at the first release stage
\begin{eqnarray}
<\zeta_{1}|\zeta>=(1+|s|^{2})\{\cos\phi^{0}\cos\phi^{1}\exp[-i(\chi_{2}^{0}-\chi_{2}^{1})]+
\sin\phi^{0}\sin\phi^{1}\exp[-i(\chi_{3}^{0}-\chi_{3}^{1})]\}\nonumber\\
\{\sin\phi^{0}\cos\phi^{1}\exp[-i(\chi_{2}^{0}-\chi_{2}^{1})]-
\cos\phi^{0}\sin\phi^{1}\exp[-i(\chi_{3}^{0}-\chi_{3}^{1})]\}
\end{eqnarray}

If in particular $\phi^{0}=0$, which means that exactly one photon has been trapped in each
coherence, then the photon coalescence probability at the first release stage is
\begin{equation}
P_{coal}(1)=\frac{1}{4}(1+|s|^{2})\sin^{2}2\phi^{1}.
\end{equation}

If the wave packets overlap ($s=1$) we obtain for $\phi^{1}=\frac{\pi}{4}$ (equal amplitudes
of the releasing control fields) that $P_{coal}(1)=\frac{1}{2}$. The same value is obtained
for the coalescence probability at the second stage. This means that the situation is
impossible in which exactly one photon is released at each stage. This is analogue of a
symmetric beam splitter with exactly one photon in each input port. A reduction of the
overlap integral leads in a continuous way to the situation in which the probability of the
latter situation grows to $\frac{1}{2}$; this is an analogue of changing the length of the
arms of the standard HOM interferometer. For example if the wave packets $f_{1,2}$ are
Gaussians with standard deviations $\delta_{1,2}$ separated by a distance $a$, the
probability of releasing a single photon at each stage (photon noncoalescence), being our
analogue of coincident registration of single photons in the usual realization,  is
\begin{equation}
P_{noncoal}=\frac{1}{2}[1-\frac{2}{\frac{\delta_{2}}{\delta_{1}}+\frac{\delta_{1}}{\delta_{2}}}
\exp(-\frac{a^{2}}{2(\delta_{1}^{2}+\delta_{2}^{2})})].
\end{equation}
This probability as a function of the distance $a$ exhibits a minimum called the Mandel dip,
reaching zero for equal packet widths (see Fig. \ref{fig2}), which corresponds to an ideal
photon coalescence. On the other hand in the case of coinciding packet centers ($a=0$) the
probability equals $(\delta_{1}-\delta_{2})^{2}/[2(\delta_{1}^{2}+\delta_{2}^{2})]$ (see
Fig. \ref{fig3}). Note that it is only for equal packets' widths that the coalescence
probability drops to zero.

The effect can also be controlled by choosing the phases of the control field. From Eq. (14)
it follows for example that for $\phi^{0}=\phi^{1}=\frac{\pi}{4}$ the coalescence
probability at the first stage is
\begin{equation}
P_{coal}(1)=\frac{1}{4}(1+|s|^{2})\sin^{2}(\chi_{2}^{0}-\chi_{2}^{1}-\chi_{3}^{0}+\chi_{3}^{1}).
\end{equation}
Thus the probability of the photons' coalescence strongly depends not only on the shape of
the wave packets and the relative amplitudes of the control fields but also, in an
oscillatory way, on the relative phases of the latter.

\section{Conclusions}
We have analyzed the propagation and storing of photons in an atomic medium in the tripod
configuration. An adiabatic evolution of the system can be described in terms of a couple of
polaritons. Using two complementary sets of control fields at the storing stage one can stop
two photons arriving at different time instants. Applying two other complementary sets of
the control fields at the release stage one retrieves two time-entangled photons. The field
operators of the outgoing photons are expressed by those of the incoming ones, depending on
the parameters of the control fields. We have thus obtained a kind of a beam splitter,
operating on a stored light, in which the input and output ports correspond to photons
arriving and leaving at different time instants. Our analogues of the transmission and
reflection coefficients as well phase relations can be regulated on demand. We have
demonstrated the effect of the Hong-Ou-Mandel - type interference and analyzed its
dependence on the shapes and positions of the trapped photon wavepackets and on the
parameters of the control fields.

\begin{acknowledgments}
The authors thank Mr. Wojciech Wasilewski for stimulating discussions. The work has been
supported by Polish budget funds allocated for the years 2005-2007 as a research grant No. 1
P03B 010 28. The subject belongs to the scientific program of the National Laboratory of AMO
Physics in Toru\'n, Poland.
\end{acknowledgments}

\newpage
\begin{figure}
\caption {\label{fig1} Level and coupling scheme of the tripod system.}
\end{figure}

\begin{figure}
\caption {\label{fig2} The probability of photon noncoalescence as a function of the
packet's separation $a$ for $\delta_{1}=\delta_{2}$ solid line, and
$\delta_{1}=3\delta_{2}$, dashed line.}
\end{figure}

\begin{figure}
\caption {\label{fig3} The probability of photon noncoalescence as a function of the ratio
$\delta_{2}\delta_{1}$ of the packets' widths for $a=0$, solid line, and $a=2\delta_{1}$,
dashed line.}
\end{figure}

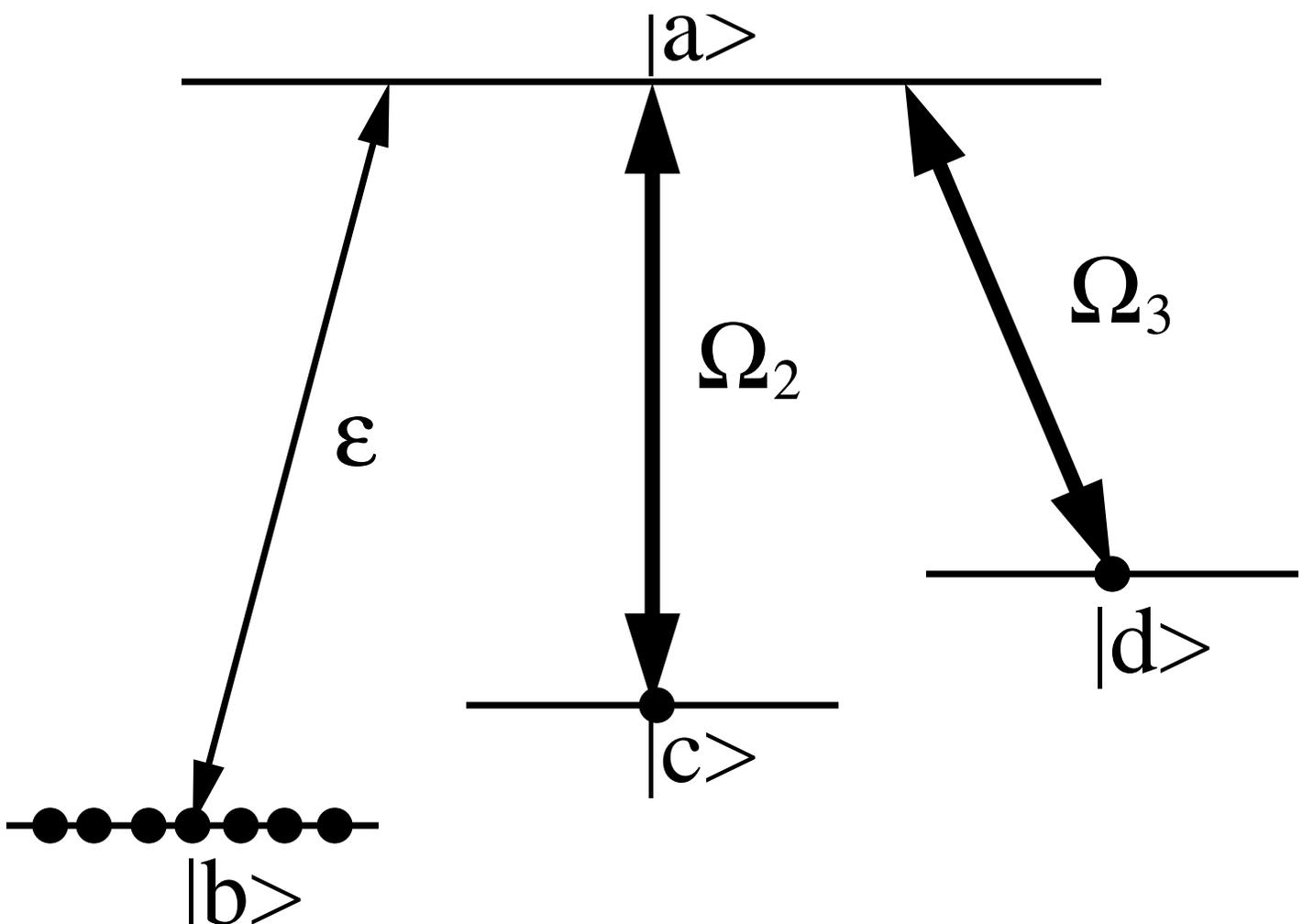
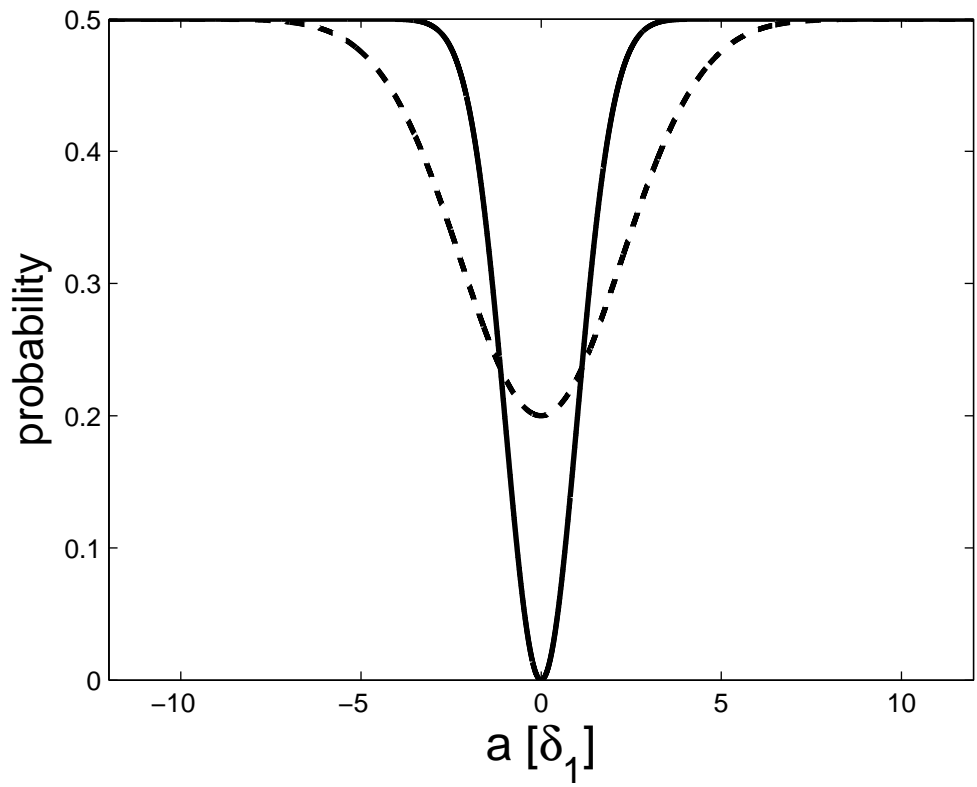
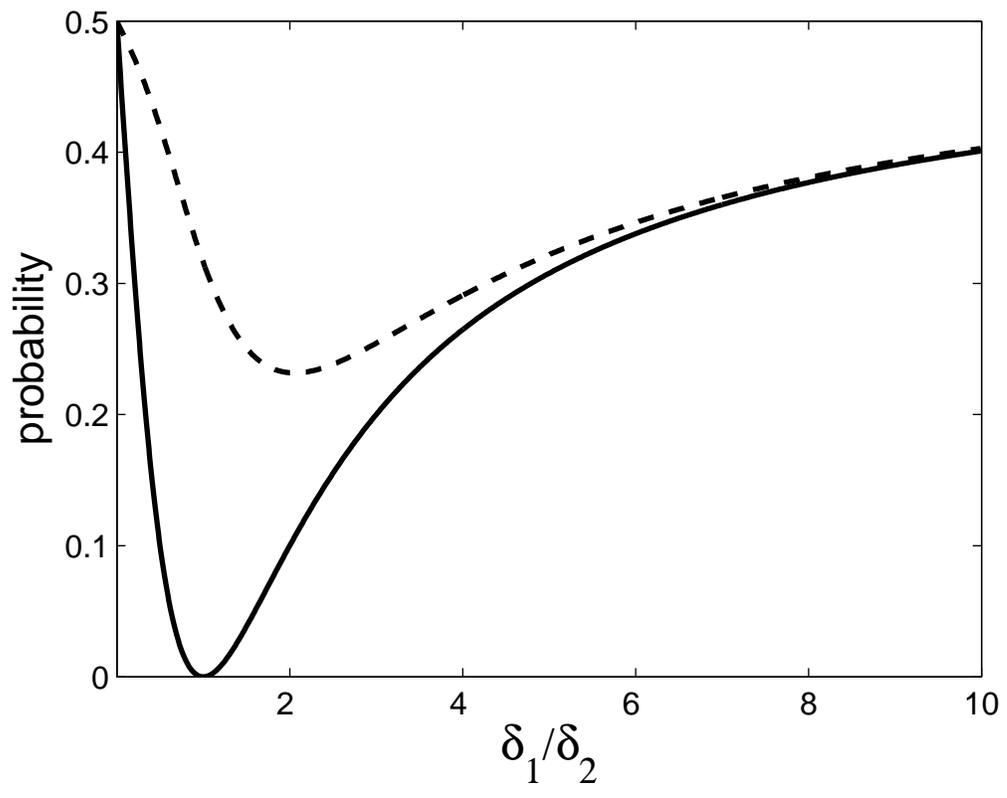
\end{document}